\def \SAIT #1 #2 {{\em Mem.\ Soc.\ Astron.\ It.\/} {\bf #1}, #2}
\def \MESS #1 #2 {{\em The Messenger\/} {\bf #1}, #2}
\def \ASTRNACH #1 #2 {{\em Astron. Nach.\/} {\bf #1}, #2}
\def \AAP #1 #2 {{\em Astron. Astrophys.\/} {\bf #1}, #2}
\def \AAL #1 #2 {{\em Astron. Astrophys. Lett.\/} {\bf #1}, L#2}
\def \AAR #1 #2 {{\em Astron. Astrophys. Rev.\/} {\bf #1}, #2}
\def \AAS #1 #2 {{\em Astron. Astrophys. Suppl. Ser.\/} {\bf #1}, #2}
\def \AJ #1 #2 {{\em Astron. J.\/} {\bf #1}, #2}
\def \ANNREV #1 #2 {{\em Ann. Rev. Astron. Astrophys.\/} {\bf #1}, #2}
\def \APJ #1 #2 {{\em Astrophys. J.\/} {\bf #1}, #2}
\def \APJL #1 #2 {{\em Astrophys.. J. Lett.\/} {\bf #1}, L#2}
\def \APJS #1 #2 {{\em Astrophys. J. Suppl.\/} {\bf #1}, #2}
\def \APSS #1 #2 {{\em Astrophys. Space Sci.\/} {\bf #1}, #2}
\def \ASR #1 #2 {{\em Adv. Space Res.\/} {\bf #1}, #2}
\def \BAIC #1 #2 {{\em Bull. Astron. Inst. Czechosl.\/} {\bf #1}, #2}
\def \JSQRT #1 #2 {{\em J. Quant. Spectrosc. Radiat. Transfer\/} {\bf #1}, #2}
\def \MN #1 #2 {{\em Mon. Not. R. Astr. Soc.\/} {\bf #1}, #2}
\def \MEM #1 #2 {{\em Mem. R. Astr. Soc.\/} {\bf #1}, #2}
\def \PLR #1 #2 {{\em Phys. Lett. Rev.\/} {\bf #1}, #2}
\def \PASJ #1 #2 {{\em Publ. Astron. Soc. Japan\/} {\bf #1}, #2}
\def \PASP #1 #2 {{\em Publ. Astr. Soc. Pacific\/} {\bf #1}, #2}
\def \NAT #1 #2 {{\em Nature\/} {\bf #1}, #2}

\documentstyle{memsait}
\input epsf.sty
\begin{opening}
\title{Coordinated RXTE and multiwavelength observations of blazars} 
\author{Rita Sambruna$^1$}
\institute{$^1$Pennsylvania State University, State College, PA 16802, USA}
\date{} % DO NOT INSERT ANY DATE HERE !!!
\end{opening}

\begin{document}

%\oddpagefooter{\sf Mem. S.A.It., Vol. ??, ??}{}{\thepage}
%\evenpagefooter{\thepage}{}{\sf Mem. S.A.It., Vol. ??, ??}
\oddpagefooter{}{}{} % LEAVE AS IT IS !
\evenpagefooter{}{}{} % LEAVE AS IT IS !
\
\bigskip

\begin{abstract}
Results from recent multiwavelength observations of blazars are
reviewed, with particular emphasis on those involving the {\it Rossi
X-ray Timing Explorer} (RXTE).  I discuss blazars' spectral energy
distributions, their correlated variability at various energies, and
the insights they offer on the physical processes in the jet. New
results on Mrk 501, PKS 2155--304, and PKS 2005--489 are highlighted.
\end{abstract} 

\section{Blazars and Their spectral energy distributions (SEDs)} 

Blazars are radio-loud Active Galactic Nuclei characterized by
polarized, highly luminous, and rapidly variable non-thermal continuum
emission (Angel \& Stockmann 1980) from a relativistic jet oriented
close to the line of sight (Blandford \& Rees 1978). As such, blazars
provide fortuitous natural laboratories to study the jet processes and
ultimately how energy is extracted from the central black hole.

The radio through gamma-ray spectral energy distributions (SEDs) of
blazars exhibit two broad humps (Figure 1). The first component peaks
at IR/optical in ``red'' blazars and at UV/X-rays in their ``blue''
counterparts, and is most likely due to synchrotron emission from
relativistic electrons in the jet (see Ulrich, Maraschi, \& Urry 1997
and references therein).  The second component extends from X-rays to
gamma-rays (GeV and TeV energies), and its origin is less well
understood. A popular scenario is inverse Compton (IC) scattering of
ambient photons, either internal (synchrotron-self Compton, SSC;
Tavecchio, Maraschi, \& Ghisellini 1998) or external to the jet
(external Compton, EC; see B\"ottcher 1999 and references therein).
In the following discussion I will assume the synchrotron and IC
scenarios, keeping in mind, however, that a possible alternative for
the production of gamma-rays is provided by the hadronic models
(proton-induced cascades; see Rachen 1999 and references therein).

Red and blue blazars are just the extrema of a continuous distribution
of SEDs. This is becoming increasingly apparent from recent multicolor
surveys (Laurent-Muehleisen et al. 1998; Perlman et al. 1998), which
find sources with intermediate spectral shapes, and trends with
bolometric luminosity were discovered (Sambruna, Maraschi, \& Urry
1996; Fossati et al. 1998).  In the more luminous red blazars the
synchrotron and IC peak frequencies are lower, the Compton dominance
(ratio of the synchrotron to IC peak luminosities) is larger, and the
luminosity of the optical emission lines/non-thermal blue bumps is
larger than in their blue counterparts (Sambruna 1997).

A possible interpretation is that the different types of blazars are
due to the different predominant electrons' cooling mechanisms
(Ghisellini et al. 1998). In a simple homogeneous scenario, the
synchrotron peak frequency $\nu_S \propto \gamma_{el}^2$, where
$\gamma_{el}$ is the electron energy determined by the competition
between acceleration and cooling. Because of the lower energy
densities, in lineless blue blazars the balance between heating and
cooling is achieved at larger $\gamma_{el}$, contrary to red blazars,
where, because of the additional external energy density, the balance
is reached at lower $\gamma_{el}$. Blue blazars are SSC-dominated,
while red blazars are EC-dominated.  While there are a few caveats to
this picture (Urry 1999), the spectral diversity of blazars' jets
cannot be explained by beaming effects {\it only} (Sambruna et
al. 1996; Georganopoulos \& Marscher 1998), but require instead a
change of physical parameters and/or a different jet environment.

%\vspace{1cm} 
\begin{figure}
\epsfysize=6cm % fix the y-dimension and scales x-dim. to y-dim.
%\epsfxsize=8cm % fix the x-dimension and scales y-dim. to x-dim.
% Feel free to do the choice you prefer but do not exceed the x-dimension
% of the text lines
\hspace{1.5cm}\epsfbox{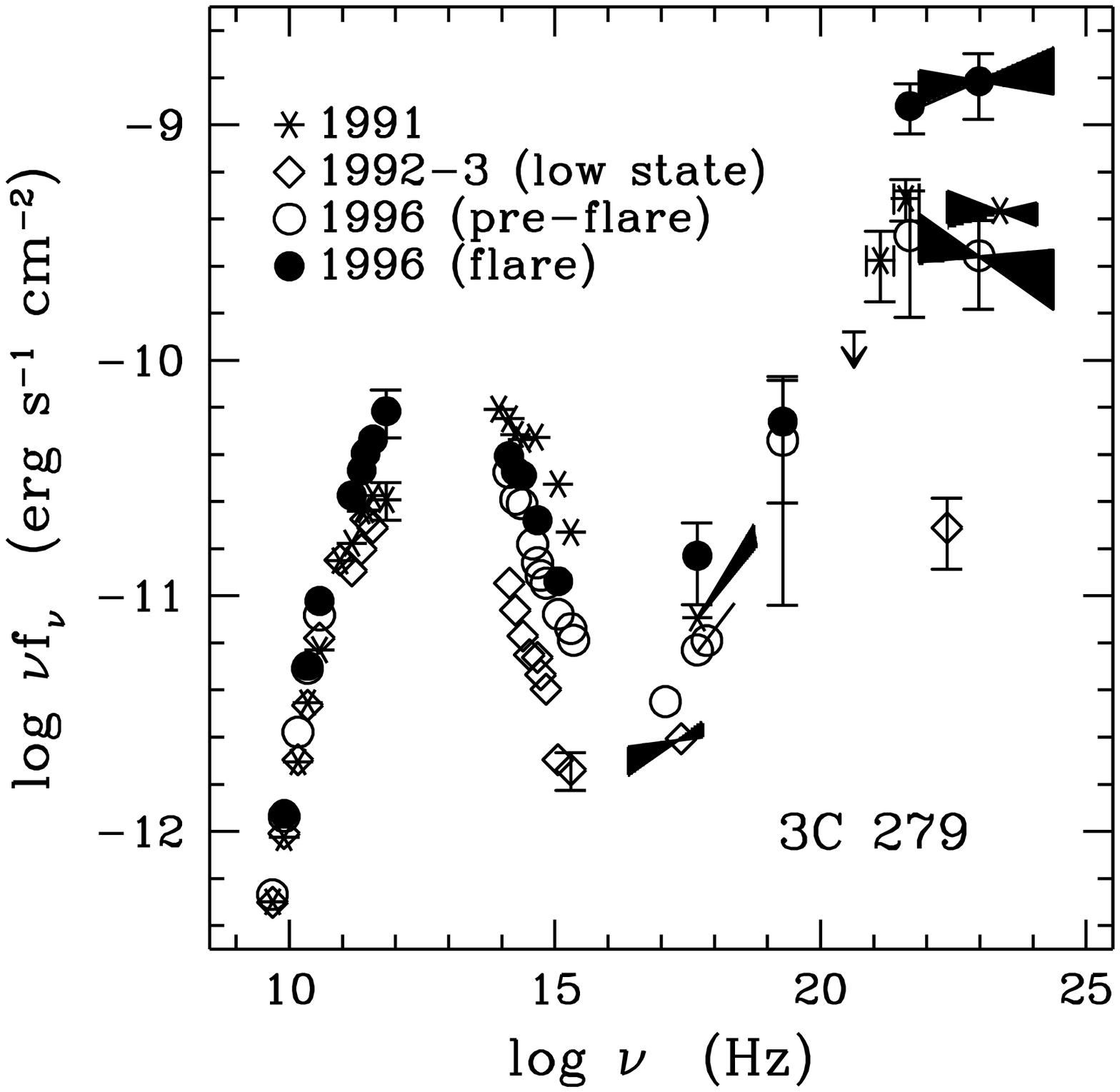}\epsfysize=6cm\epsfbox{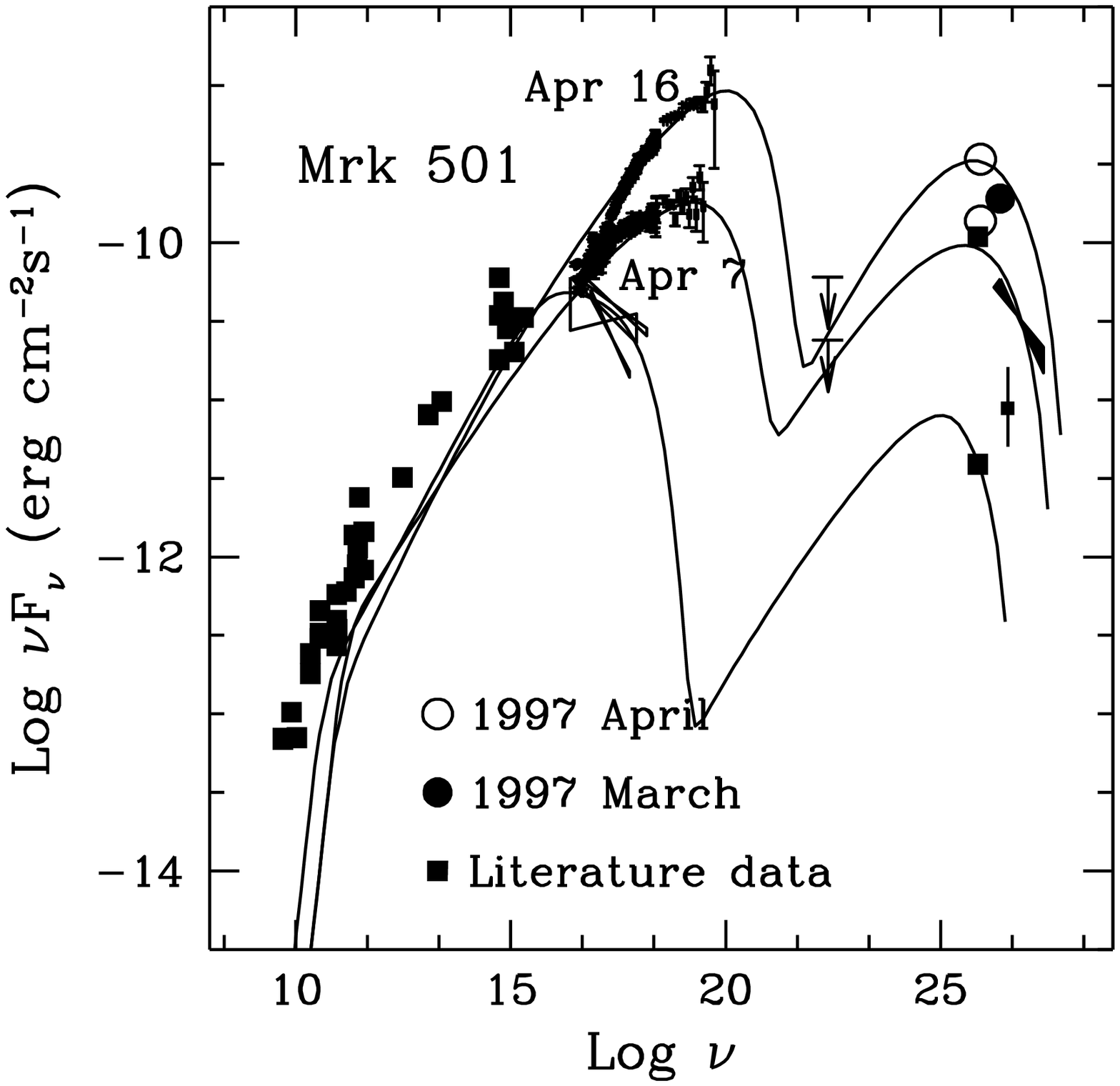} 
%for centering: act on hspace argument 
\caption[h]{Spectral energy distributions (SEDs) of 
3C279 [{\it (a), Left]} and Mrk 501 [{\it (b), Right]}. Data are from
Maraschi et al. (1994), Wehrle et al. (1998), and Pian et
al. (1998). Blazars' SEDs typically have two broad humps, the first
peaking anywhere from IR/optical (in red blazars like 3C279) to hard
X-rays (in blue blazars like Mrk 501) and due to synchrotron emission
from a relativistic jet.  The second component, extending to
gamma-rays, is less well understood. A popular explanation is inverse
Compton scattering of ambient seed photons off the jet's electrons.}
\end{figure}

\section{Correlated multiwavelength variability: Testing the 
blazar paradigm} 

Correlated multiwavelength variability provides a way to test the
cooling paradigm since the various synchrotron and IC models make
different predictions for the relative flare amplitudes and shape, and
the time lags.  First, since the same population of electrons is
responsible for emitting both spectral components (in a homogeneous
scenario), correlated variability of the fluxes at the low- and
high-energy peaks with no lags is expected (Ghisellini \& Maraschi
1996).  Second, if the flare is caused by a change of the electron
density and/or seed photons, for a fixed beaming factor $\delta$ the
relative amplitudes of the flares at the synchrotron and IC peaks obey
simple and yet precise relationships (Ghisellini \& Maraschi 1996; see
however B\"ottcher 1999). Third, the rise and decay times of the
gamma-ray flux are a sensitive function of the external gas opacity
and geometry in the EC models (B\"ottcher \& Dermer 1998). Fourth, the
rise and decay times of the synchrotron flux depend on a few source
typical timescales (Chiaberge \& Ghisellini 1999). Fifth, spectral
variability accompanying the synchrotron flare (in X-rays for blue
blazars, in optical for red blazars) is a strong diagnostic of the
electron acceleration versus cooling processes (Kirk, Riegler, \&
Mastichiadis 1998). When cooling dominates, time lags between the
shorter and longer synchrotron wavelengths provide an estimate of the
magnetic field $B$ (in Gauss) of the source via $t_{lag} \sim t_{cool}
\propto E^{-0.5} \delta^{-0.5} B^{-1.5}$ (Takahashi et al. 1996; 
Urry et al. 1997). 

\noindent {\bf The role of RXTE.} 
With its wide energy band coverage (2--250 keV), RXTE plays a crucial
role in monitoring campaigns of blazars, since it probes the region
where the synchrotron and Compton component overlap in the SEDs
(Figure 1), allowing us to quantify their relative importance in the
different sources. Its high time resolution and good sensitivity are
ideal to detect the smallest X-ray variability timescales, study the
lags between the harder and softer X-rays, and to follow the particle
spectral evolution down to timescales of a few hours or less, pinning
down the microphysics of blazars' jets.

\subsection{Results for red blazars} 

One of the best monitored red blazars is 3C279. From the simultaneous
or contemporaneous SEDs in Figure 1a, it is apparent that the largest
variations are observed above the synchrotron peak in IR/optical (not
well defined) and the Compton peak at GeV energies, supporting the
synchrotron and IC models.  The GeV amplitude is roughly the square of
the optical flux during the earlier campaigns, supporting an SSC
interpretation (Maraschi et al. 1994) or a change of $\delta$ in the
EC models, while in 1996 large variations were recorded at gamma-rays
but not at lower energies (Wehrle et al. 1998). During the latter
campaign, the rapid decay of the GeV flare (Figure 2a) favors an EC
model (B\"ottcher \& Dermer 1998; Wehrle et al. 1998).  Note in Figure
2a the good correlation, within one day, of the EGRET and RXTE flares,
which provides the first evidence that the gamma-rays and X-rays are
cospatial (Wehrle et al. 1998).

%\vspace{1cm} 
\begin{figure}
\epsfysize=7.5cm % fix the y-dimension and scales x-dim. to y-dim.
%\epsfxsize=8cm % fix the x-dimension and scales y-dim. to x-dim.
% Feel free to do the choice you prefer but do not exceed the x-dimension
% of the text lines
\hspace{0.4cm}\epsfbox{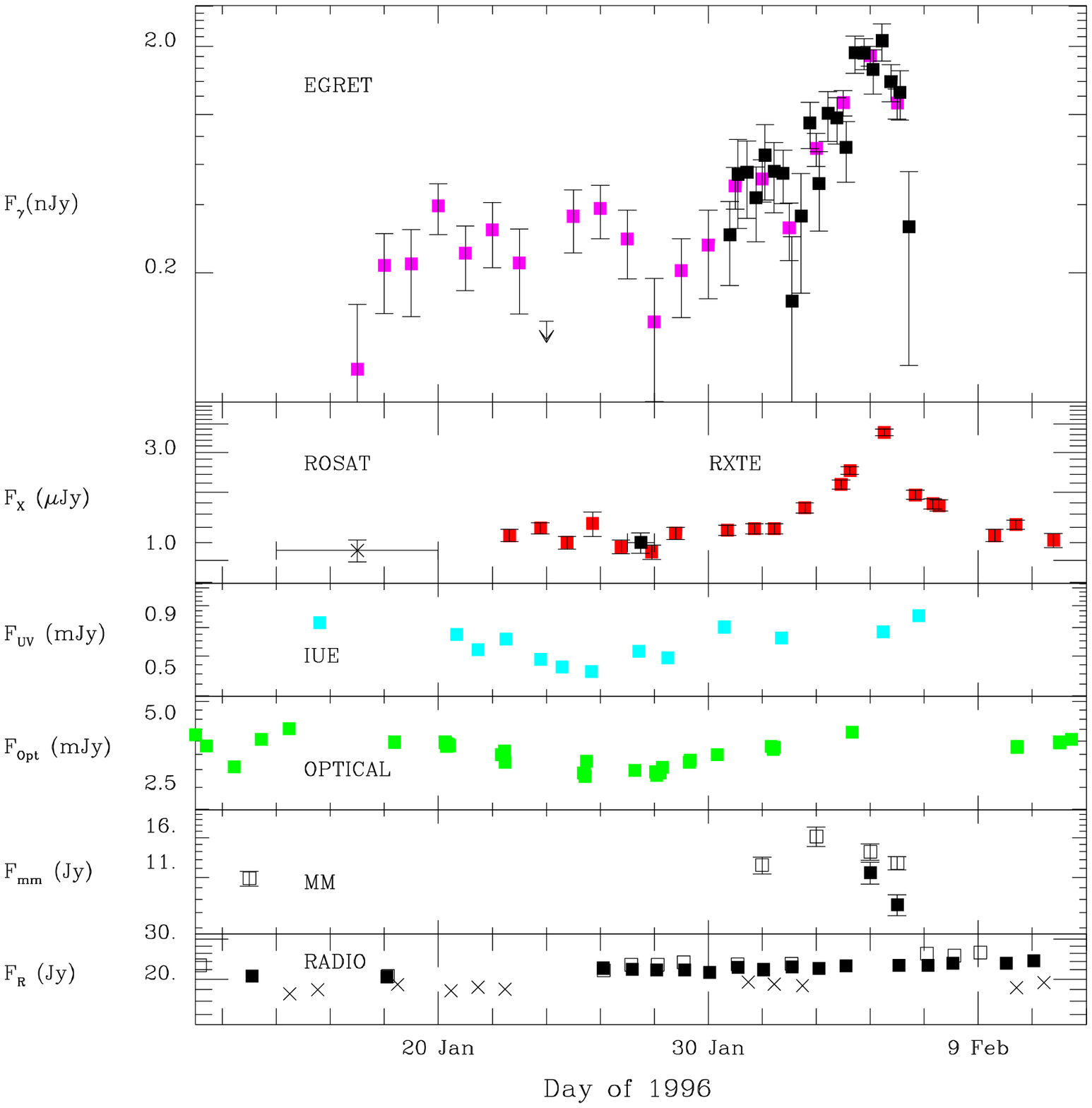}\epsfysize=7.5cm\epsfbox{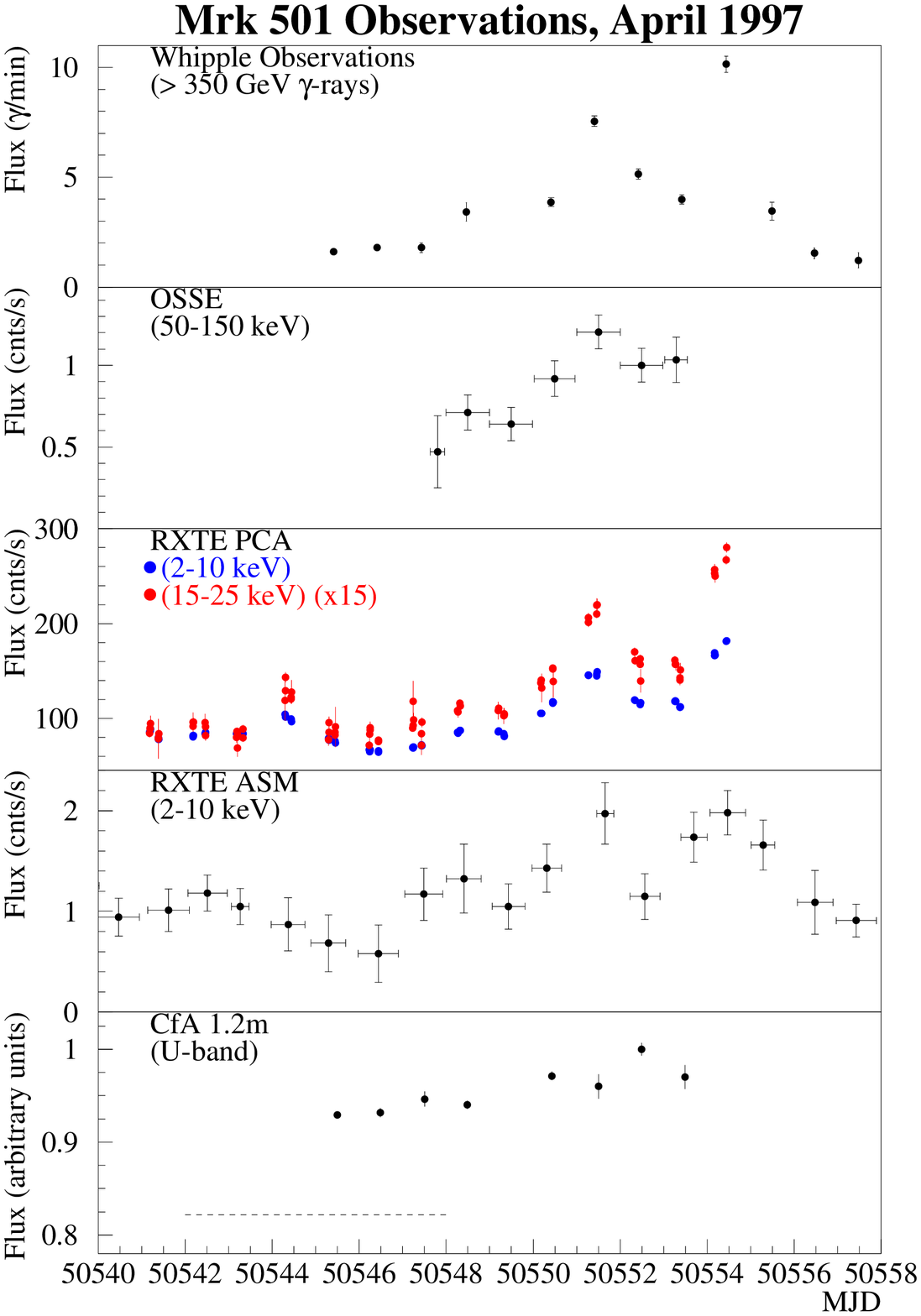} 
%for centering: act on hspace argument 
\caption[h]{Multiwavelength light curves of 3C279 [{\it
(a), Left]} and Mrk 501 [{\it (b), Right]}. Data from Wehrle et
al. (1998) and Catanese (1999).}
\end{figure} 

Another candidate for future gamma-ray monitorings is BL Lac
itself. In 1997 July it underwent a strong flare at GeV and optical
energies (Bloom et al. 1997). The gamma-ray light curve shows a strong
flare possibly anticipating the optical by up to 0.5 days; however,
the poor sampling does not allow firmer conclusions. Contemporaneous
RXTE observations showed a harder X-ray continuum (Madejski et
al. 1999) than in previous ASCA measurements. The SED during the
outburst is best modeled by the SSC model from radio to X-rays, while
an EC contribution is required above a few MeV (Sambruna et
al. 1999a). A similar mix of SSC and EC is also required to fit the
SEDs of PKS 0528+134 (Sambruna et al. 1997; Mukherjee et al. 1998;
Ghisellini et al. 1999). 

\subsection{Results for blue blazars} 

Mrk 501, one of the two brightest TeV blazars, attracted much
attention in 1997 April when it underwent a spectacular flare at TeV
energies (Catanese et al. 1997; Aharonian et al. 1999; Djannati-Atai
et al. 1999). This was correlated to a similarly-structured X-ray
flare observed with RXTE (Figure 2b), with no delay larger than one
day (Krawczynski et al. 1999).  These results are consistent with an
SSC scenario where the most energetic electrons are responsible for
both the hard X-rays (via synchrotron) and the TeV (via IC).

Figure 1b shows the SEDs of Mrk 501 during the 1997 April TeV
activity, compared to the ``quiescent'' SED from the literature.  An
unusually flat (photon index, $\Gamma_X \sim 1.8$) X-ray continuum was
measured by SAX and RXTE during the TeV flare ( Pian et al. 1998;
Krawczynski et al. 1999), implying a shift of the synchrotron peak by
more than two orders of magnitude. This almost certainly reflects a
large increase of the electron energy (Pian et al. 1998), or the
injection of a new electron population on top a quiescent one (Kataoka
et al. 1999a). Later RXTE observations in 1997 July found the source
still in a high and hard X-ray state (Lamer \& Wagner 1998),
indicating a persistent energizing mechanism.

An interesting new behavior was observed during our latest 2-week
RXTE-HEGRA monitoring of Mrk 501 in 1998 June (Sambruna et al. 1999b),
when 100\% overlap between the X-rays and TeV light curves was
achieved (Figure 3a). A strong short-lived ($\sim$ two days) TeV flare
was detected, correlated to a flare in the very hard (20--50 keV)
X-rays, with the softer X-rays being delayed by up to one day. As in
1997, large X-ray spectral variations are observed, with the X-ray
continuum flattening to $\Gamma_X=1.9$ at the peak of the TeV flare,
implying a similar shift to $\ge$ 50 keV of the synchrotron peak
(Figure 3b). However, while in 1997 the TeV spectrum hardened during
the flare (Djannati-Atai et al. 1997), as it did in the X-rays, we did
not observe significant variability in the TeV hardness ratios during
the flare (Figure 3a, panel (e)); instead the spectrum softens 1-2
days later. The correspondence between the X-ray and TeV spectra is no
longer present during the 1998 June flare.

%No shifts of the synchrotron peak were ever reported for Mrk 421, the
%other bright TeV blazar (while they were observed in other blue
%blazars). In this source, the TeV and hard X-rays are well correlated
%down to one hour (Maraschi et al. 1999). 

%\vspace{1cm} 
\begin{figure}
\epsfysize=8cm % fix the y-dimension and scales x-dim. to y-dim.
%\epsfxsize=8cm % fix the x-dimension and scales y-dim. to x-dim.
% Feel free to do the choice you prefer but do not exceed the x-dimension
% of the text lines
\hspace{0.1cm}\epsfbox{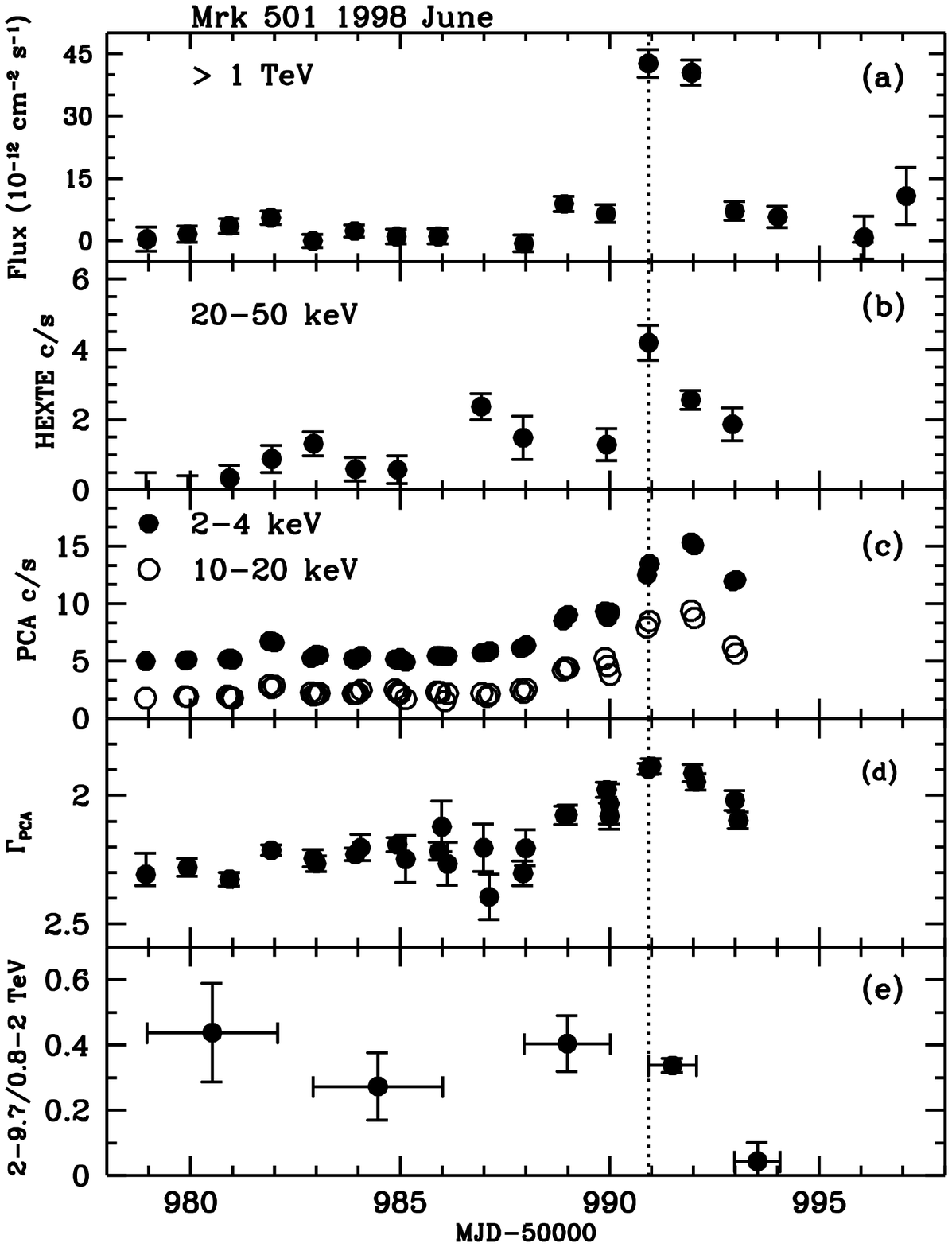}\epsfysize=5cm\epsfbox{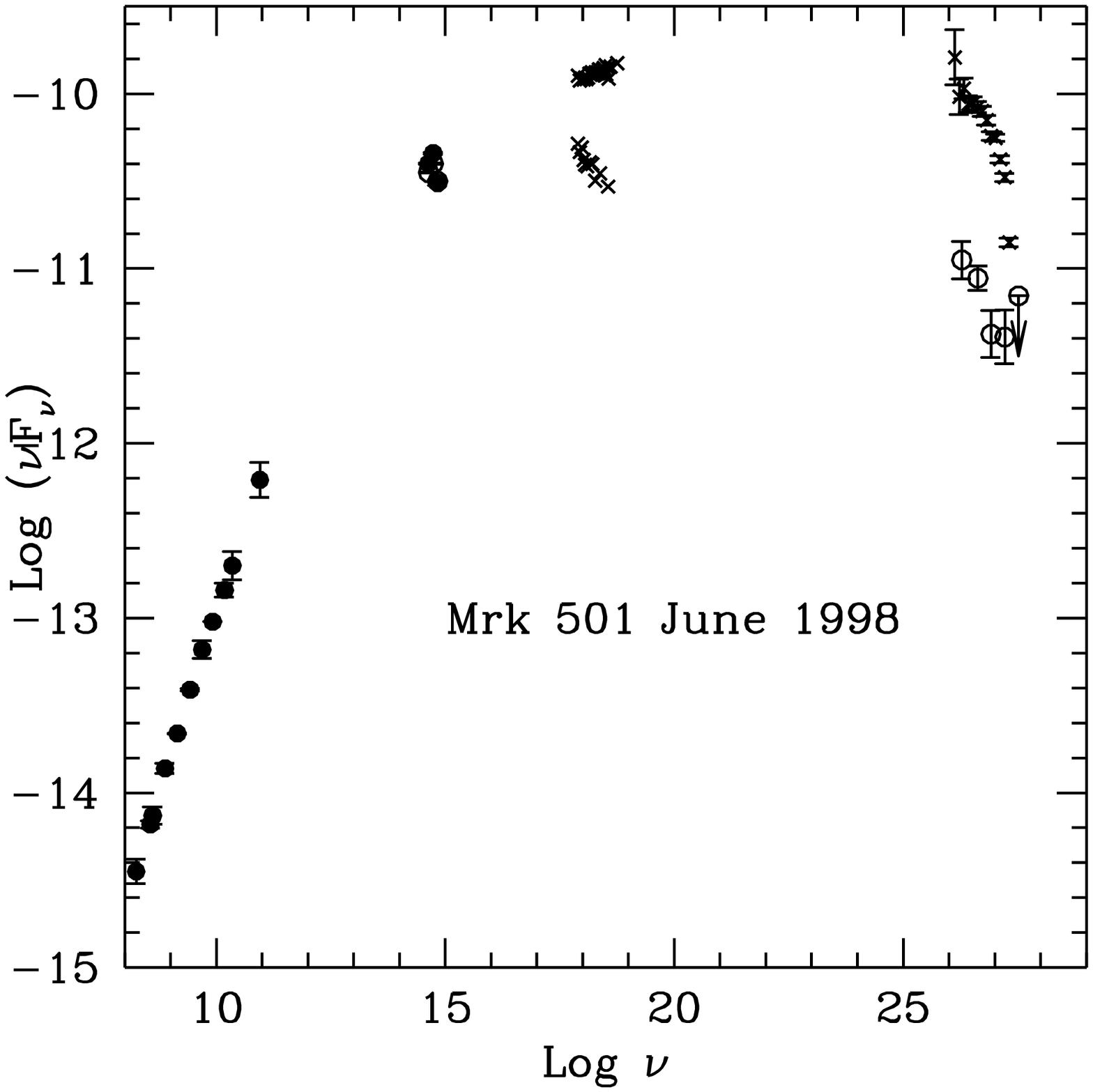}
%for centering: act on hspace argument 
\caption[h]{RXTE and HEGRA monitoring of Mrk 501 in 1998 June (Sambruna et
al. 1999b). {\it (a), Left:} Multifrequency light curves and spectral
variations at both X-rays (panel (d)) and TeV (panel (e)).  The TeV
flare is well correlated with the harder X-rays, while the softer
X-rays appear to peak up to one day later. While the X-ray spectrum
hardens during the TeV flare, the TeV hardness ratios stay constant,
and decrease one day later.  {\it (b), Right:} SEDs of Mrk 501 during
the high (filled symbols) and low (open symbols) states in 1998
June. Note the shift of the synchrotron peak above 50 keV occurring on
about a few days, similar to the more persistent 1997 April flare, and
the curved TeV spectra.}
\end{figure}

\section{Acceleration and cooling in blue blazars} 

X-ray monitorings of blue blazars are a powerful diagnostic of
physical processes occurring in these sources. This is because in
these objects the X-rays are the high-energy tail of the synchrotron
component where rapid and complex flux and spectral variability is
expected depending on the balance between escape, acceleration, and
cooling of the emitting particles (Kirk et al. 1998). 

An ideal target for X-ray monitorings is PKS 2155--304, one of the
brightest X-ray blazars (Treves et al. 1989; Sembay et al. 1993; Pesce
et al. 1998).  Interest in this source was recently revived due to a
TeV detection (Chadwick et al. 1999) during a high X-ray state
(Chiappetti et al. 1999). ASCA and SAX observations detected strong
X-ray variability, with the softer energies lagging the harder
energies by one hour or less (Chiappetti et al. 1999; Kataoka et
al. 1999b; Zhang et al. 1999), and are consistent with a model where
the electron cooling dominates the flares. This implies magnetic
fields of $B
\sim 0.1-0.2$ Gauss (for $\delta \sim 10$), similar to Mrk 421
(Takahashi et al. 1996). 

A new mode of variability was discovered during our RXTE monitoring of
PKS 2155--304 in 1996 May, as part of a larger multifrequency campaign
(Sambruna et al. 1999c). The sampling in the X-rays was excellent
(Figure 4a), and complex flux variations were observed, with short,
symmetric flares superposed to a longer baseline trend. Inspection of
the hardness ratios (the ratio of the counts in 6--20 keV over the
counts in 2--6 keV) versus flux shows that different flares (separated
by vertical dashed lines in Figure 4a) exhibit hysteresis loops of
opposite signs, both in a ``clockwise'' and ``anti-clockwise'' sense
(labeled as C and A in Figure 4a, respectively).  Applying a
correlation analysis to each flare separately, we find that the C
loops corresponds to a soft lag (softer energies lagging) and the A
loop corresponds to a hard lag (harder energies lagging), of the order
of a few hours in both cases.  Two examples are shown in Figure 5b for
the May 18.5--20.2 and 24.2--26.9 flares, respectively.

We interpreted the data using the acceleration model of Kirk et
al. (1998). Here loops/lags of both signs are expected depending on
how fast the electrons are accelerated compared to their cooling time,
$t_{cool}$. If the acceleration is instantaneous (i.e., $t_{acc} <<
t_{cool}$), cooling dominates variability and, because of its energy
dependence, the harder energies are emitted first, with C loops and
soft lags. If instead the acceleration is slower ($t_{acc} \sim
t_{cool}$), the electrons need to work their way up in energy and the
softer energies are emitted first, with A loops and hard lags
predicted. A close agreement between the RXTE light curve and the
model is found by increasing $t_{acc}$ by a factor 100 going from a C to
an A loop, when $t_{acc}$ becomes similar to the duration of the
flare, and by steepening the electron energy distribution (see
Sambruna et al. 1999d for details). Thus we reach the important
conclusion that we are indeed observing electron acceleration, and
together with cooling this is responsible for the observed X-ray
variability properties of PKS 2155--304.

The complex spectral behavior of PKS 2155--304 in 1996 May is in
contrast to the remarkable simplicity observed in earlier X-ray
observations of this source (e.g., Kataoka et al. 1999b) and of other
objects.  Figure 5 summarizes two epochs of RXTE monitoring of another
bright blue blazar, PKS 2005--489, which has an SED very similar to
PKS 2155--304 and is a TeV candidate (Sambruna et al. 1995). During
1998 September, our RXTE monitoring detected a general trend of flux
increase of 30\% amplitude in 3.5 days (Figure 5a). Despite gaps in
the sampling, it is clear that the variability at harder energies is
faster than at softer energies, consistent with cooling dominating the
flares. This is confirmed by the analysis of the hardness ratios
versus flux, where only clockwise loops are observed. A consistent
behavior was also observed one month later (Figure 5b) during a much
larger, longer-lasting flare, when spectral variations occurred on
timescales of a few hours (Perlman et al. 1999).
 
%\vspace{1cm} 
\begin{figure}
\epsfysize=7cm % fix the y-dimension and scales x-dim. to y-dim.
%\epsfxsize=8cm % fix the x-dimension and scales y-dim. to x-dim.
% Feel free to do the choice you prefer but do not exceed the x-dimension
% of the text lines
\hspace{0.01cm}\epsfbox{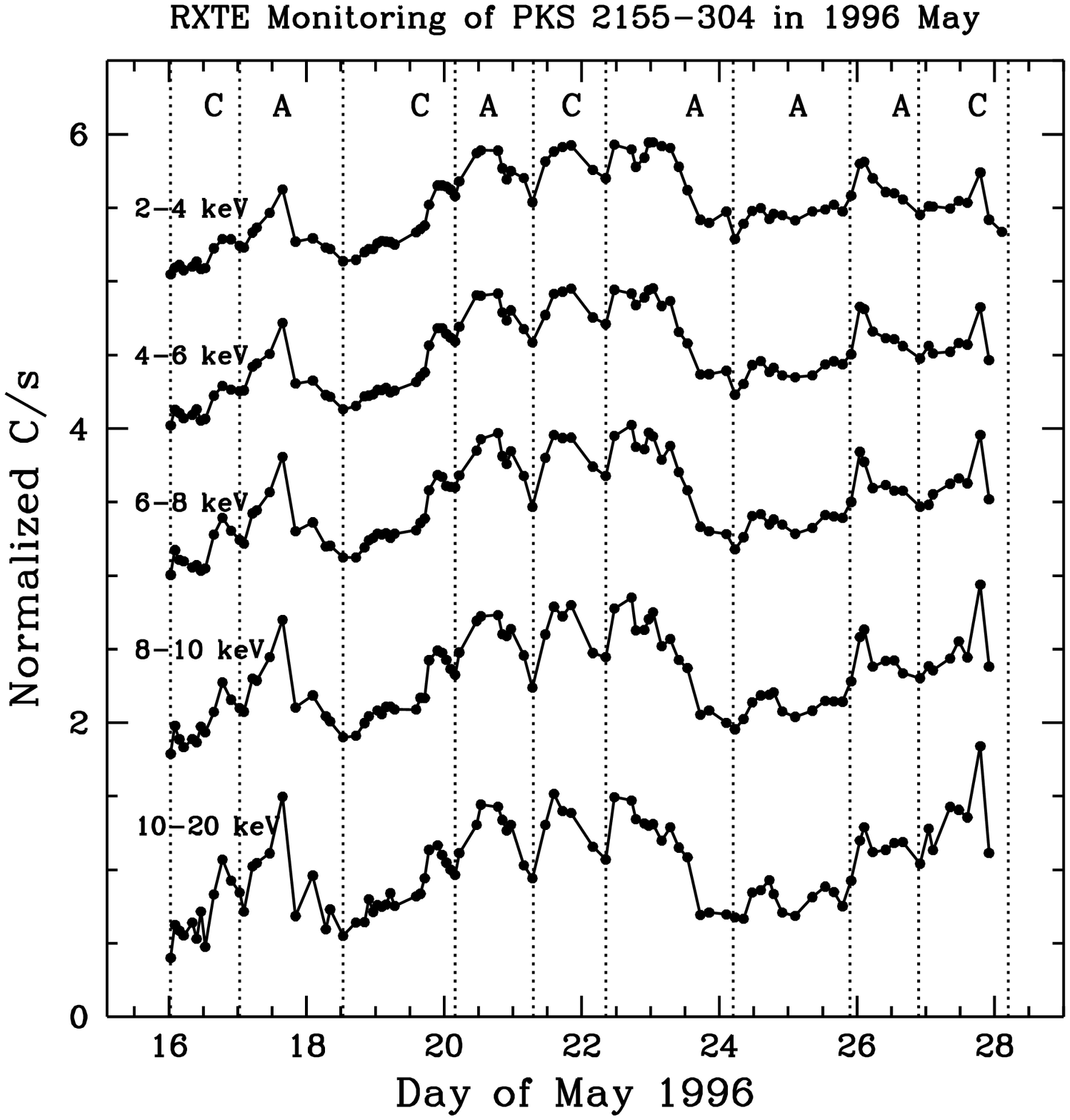}\epsfysize=6cm\epsfbox{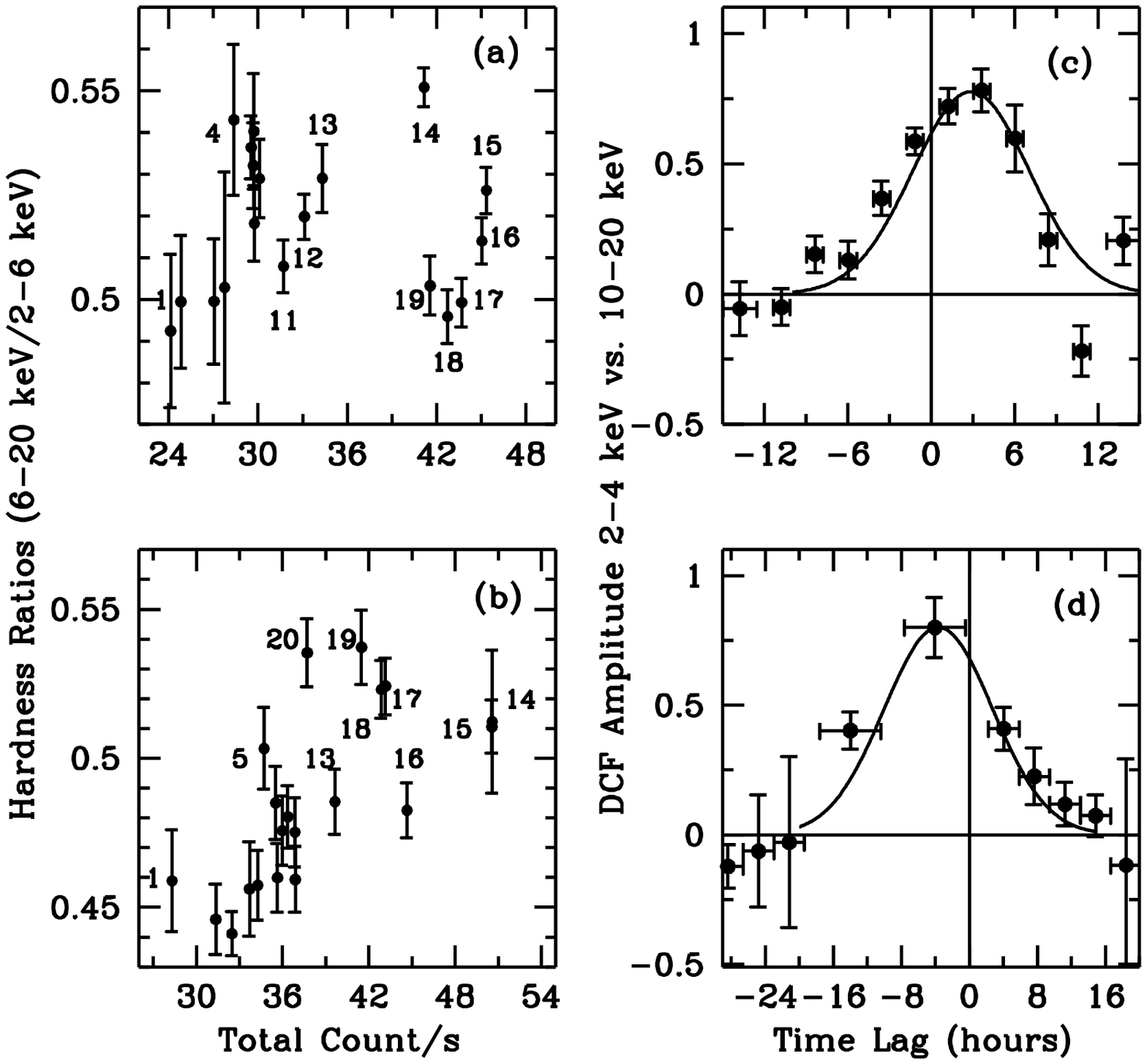}
%for centering: act on hspace argument 
\caption[h]{RXTE intensive monitoring of PKS 2155--304 in 1996 May 
(Sambruna et al. 1999c). {\it (a), Left:} Energy-dependent X-ray
light curves, normalized to their average intensity and arbitrarily
shifted. The vertical dashed lines mark portions of the light curves
characterized by ``clockwise'' (C) or ``anti-clockwise'' (A)
hysteresis loops in the hardness ratio versus intensity diagrams.
{\it (b), Middle:} Two examples of C (upper panel) and A (lower panel)
loops for the May 18.5--20.2 and 24.2--26.9 flares, respectively. {\it
(b), Right:} Discrete Correlation Function (Edelson
\& Krolik 1988) applied to the same flares. Soft and hard lags are
detected for the clockwise and anti-clockwise loops, respectively, of
$\sim$ a few hours. This complex spectral behavior is a powerful
diagnostic of the acceleration and cooling processes in the jet.}
\end{figure}

%\vspace{1cm} 
\begin{figure}
\epsfysize=7cm % fix the y-dimension and scales x-dim. to y-dim.
%\epsfxsize=8cm % fix the x-dimension and scales y-dim. to x-dim.
% Feel free to do the choice you prefer but do not exceed the x-dimension
% of the text lines
\hspace{0.5cm}\epsfbox{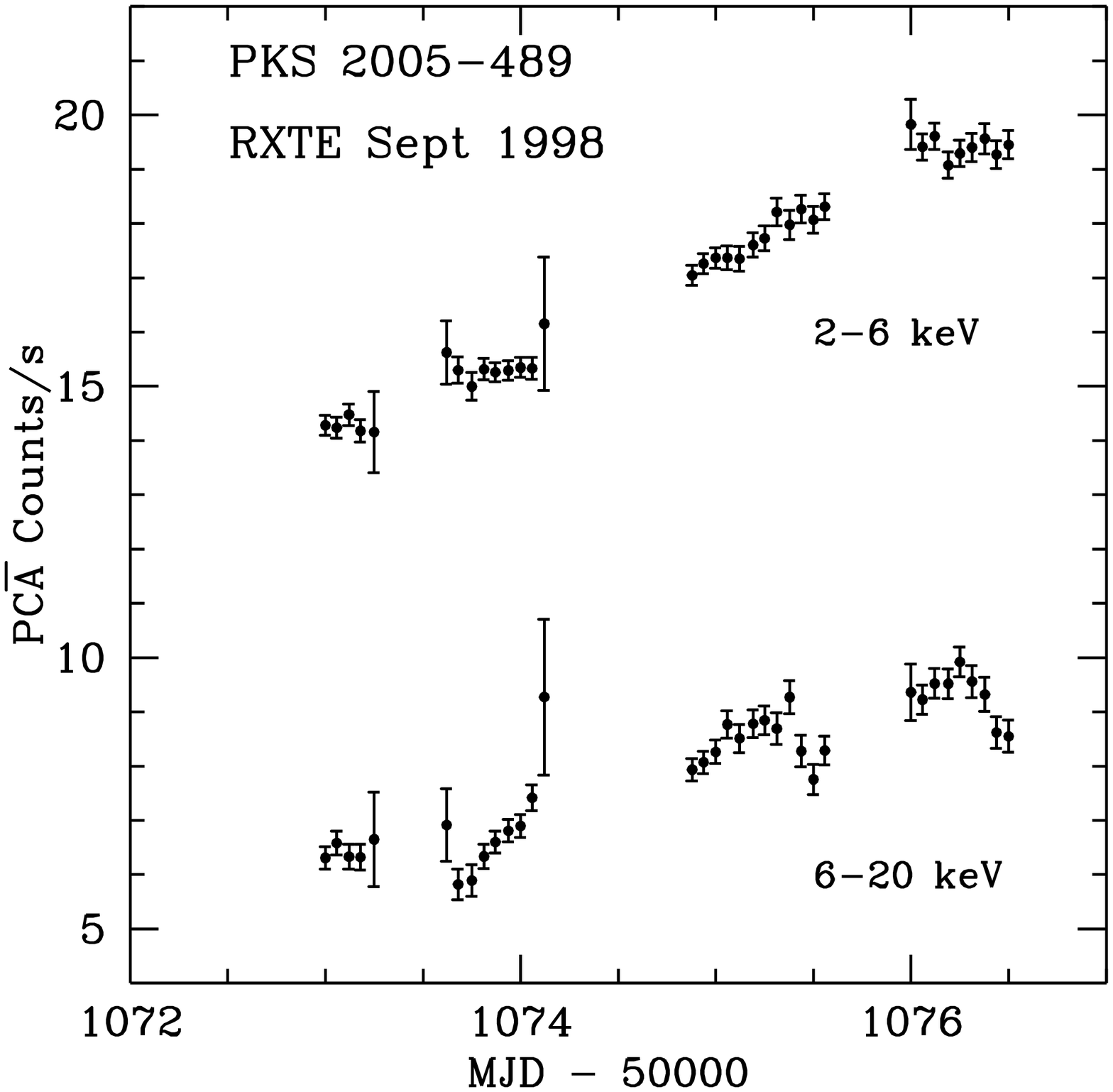}\epsfysize=7cm\epsfbox{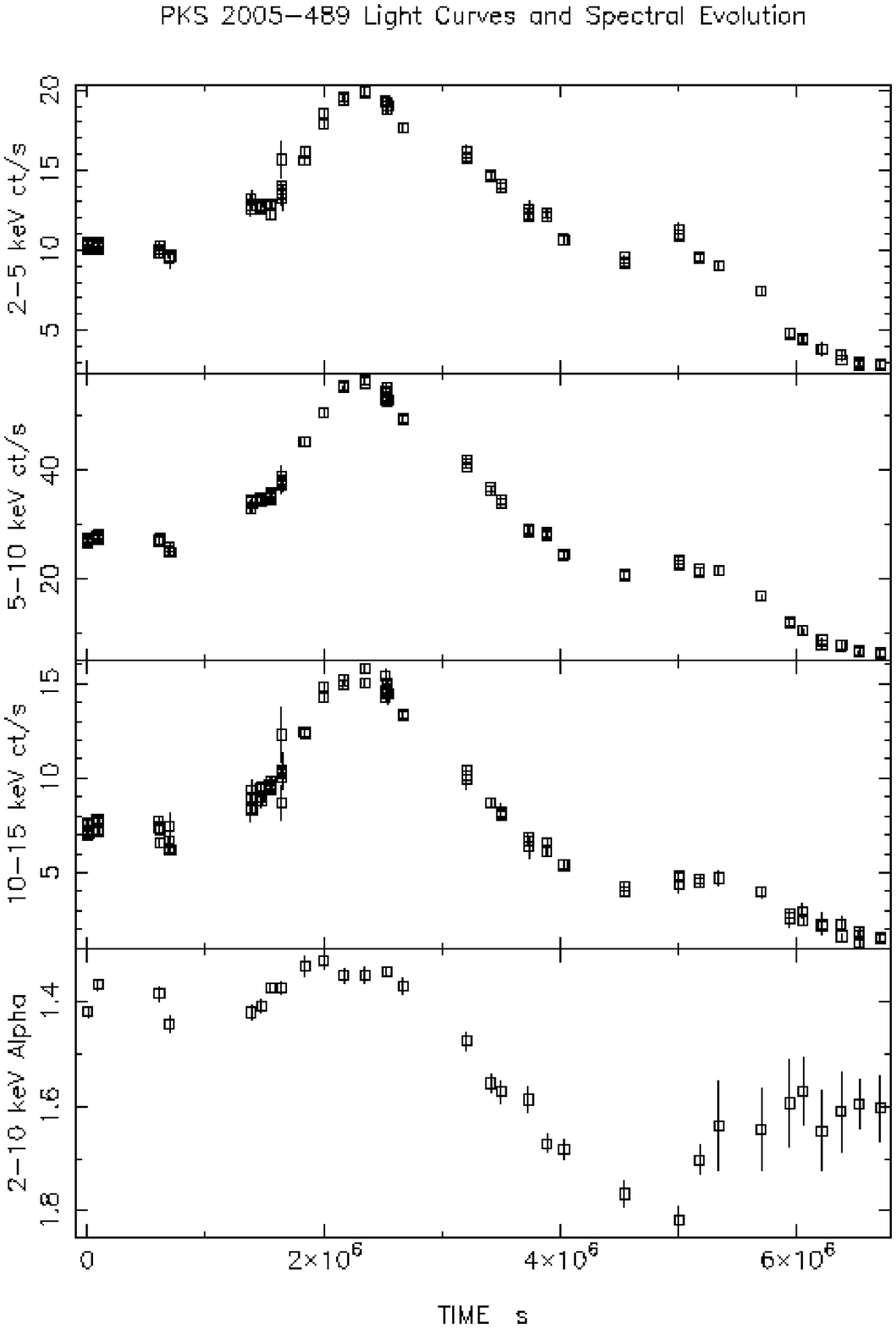}
%for centering: act on hspace argument 
\caption[h]{RXTE monitoring of PKS 2005--489 in 1998 September 
{\it [(a), Left]}. Despite the gaps, energy-dependent variability is
apparent, with a general flux increase of 30\% or more amplitude.  The
harder energies vary before the softer ones, consistent with cooling
dominating the flux variability.  {\it [(b), Right]:} One month later,
PKS 2005--489 underwent a strong, long-lasting X-ray flare which was
well sampled by RXTE (Perlman et al. 1999).}
\end{figure}

\section{Summary and Future work} 

Recent multiwavelength campaigns of blazars expanded the current
available database, from which we are learning important new lessons.
Detailed modeling of the SEDs of bright gamma-ray blazars of the red
and blue types tend to support the current cooling paradigm, where the
different blazars flavors are related to the predominant cooling
mechanisms of the electrons at the higher energies (EC in more
luminous sources, SSC in lower-luminosity ones).  However, several
observational biases could be present which can be addressed by future
larger statistical samples, especially in gamma-rays.  In particular,
it will be important to expand the sample of TeV blazars, which
currently includes only a handful (5) of sources, with only two bright
enough to allow detailed spectral and timing analysis.

A promising diagnostic for the origin of the seed photons for the IC
process is the shape of the gamma-ray flare. This awaits well-sampled
gamma-ray light curves, which will be afforded by the next
higher-sensitivities missions (GLAST, AGILE in GeV and HESS, VERITAS,
MAGIC, CANGAROO II in TeV). Broader-band, higher quality gamma-ray
spectra will also be available, allowing a better location of the IC
peak, a more precise measure of the spectral shape at gamma-rays, and
its variability.  More correlated X-ray/TeV monitorings are necessary,
in which RXTE and SAX have crucial roles, to add to the current
knowledge of the variability modes. 

Finally, the current data show that acceleration and cooling are the
dominant physical mechanisms responsible for the observed variability
properties from blazars' jets. We are starting to study these
processes well in the X-rays for blue blazars, where RXTE has the
potential to determine the shortest X-ray variability timescales and
lags, probing into even more detail the jets' microphysics. It would
also be interesting to perform similar studies in the optical for red
blazars, to compare the nature of the acceleration and cooling in the
two subclasses.

\acknowledgements
This work was supported by NASA contract NAS--38252 and NASA grant
NAG5--7276. I thank Laura Maraschi for a critical reading of the
manuscript, Felix Aharonian and the HEGRA team for allowing me to
report the 1998 TeV data of Mrk 501, and Lester Chou for help with the
RXTE data reduction.

% References. We avoided using the \bibitem commmand since we found it is
% somewhat platform-dependent. We also avoided using the \cite{keyword}
% command since we found it cumbersome. However, if you are an expert 
% LateX user you may use the various LateX tools for the references 
% provided they give the same printout formats of the examples given here.

\end{document}